# Partial Shading Detection and Smooth Maximum Power Point Tracking of PV Arrays under PSC

Mohammad Amin Ghasemi*, Hossein Mohammadian Forushani*, Mostafa Parniani* *Senior Member, IEEE*

*Abstract*— One of the most important issues in the operation of a photovoltaic (PV) system is extracting maximum power from the PV array, especially in partial shading condition (PSC). Under PSC, P-V characteristic of PV arrays will have multiple peak points, only one of which is global maximum. Conventional maximum power point tracking (MPPT) methods are not able to extract maximum power in this condition. In this paper, a novel two-stage MPPT method is presented to overcome this drawback. In the first stage, a method is proposed to determine the occurrence of PSC, and in the second stage, using a new algorithm that is based on ramp change of the duty cycle and continuous sampling from the P-V characteristic of the array, global maximum power point of array is reached. P&O algorithm is then re-activated to trace small changes of the new MPP. Open loop operation of the proposed method makes its implementation cheap and simple. The method is robust in the face of changing environmental conditions and array characteristics, and has minimum negative impact on the connected power system. Simulations in Matlab/Simulink and experimental results validate the performance of the proposed methods.

***Index Terms*— DC/DC converter, maximum power point tracking (MPPT), Partial shading condition, Photovoltaic power generation system.**

## I. Introduction

There has been increasing interest in photovoltaic (PV) systems as a renewable energy source in recent years. PV systems can be operated as grid-connected or stand-alone structures. The main element of a PV system is PV array that is a set of PV modules connected in series and parallel. In a PV array, voltage and current have a nonlinear relation, and only in one operating voltage, maximum power is generated. Therefore, extracting maximum power from a PV system in all operation conditions is the main target of its control. To date, numerous maximum power point tracking (MPPT) techniques have been presented and implemented. Some of the conventional and most popular ones are perturb and observe (P&O), incremental conductance (IC), and short circuit current and open circuit voltage. Some techniques are also presented based on artificial intelligence, such as fuzzy logic and neural network, but have more computation load [1].

A condition in which the entire modules of an array do not receive the same solar irradiance is called partial shading condition (PSC). PSCs are inevitable especially in solar systems installed in urban areas and in areas where low moving clouds are common [2]. If the control system cannot detect and react to this situation, the PV system will be diverted from the optimal operation mode. In PSC, because of bypass diodes in parallel with each module, P-V characteristic of the array has multiple peak points [3]. Conventional MPPT techniques are unable to identify the global maximum power point (GMPP) in PSC, and usually track local peaks. Therefore, developing new MPPT techniques for dealing with PSC is necessary.

In recent years, many techniques have been presented for MPPT under PSC [4-21]. Most of these techniques consist of two steps to attain GMPP. In the first step, the neighborhood of GMPP is determined, and in the second step that usually uses conventional MPPT methods such as P&O, the exact GMPP is obtained. In [4], after PSC detection, by moving on the load line that is based on short circuit current and open circuit voltage of the array, the operating point moves to the vicinity of the GMPP, and in the second step, the operating point converges to it. One can easily show that this technique is unable to track the GMPP in all PSCs [5]. The proposed method in [6] is basically a P&O algorithm that its voltage step sizes are determined based on dividing rectangles method. This technique does not guarantee reaching the GMPP. A neural network training for different PSCs is presented in [7], which is system dependent and needs measurement of solar irradiance level and temperature. Ref. [8] uses a multilevel converter and a new control algorithm to overcome the PSC problem. A novel distributed maximum power point tracking (DMPPT) is proposed in [9] wherein the current of each module is compensated by regulating its voltage at the respective MPP value by connecting a fly-back dc-dc converter in parallel with each module. The proposed MPPT in [10] uses a controllable current transformer (CCT) disposed at the terminal of each PV module, permitting compatible current in the series path of a PV string. The CCT output current can be regulated using a dependent current source according to the MPPT algorithm. Although accuracy of these methods is high and they decrease the effect of PS on the array power, their implementation is expensive.

In [11], when the PV power suddenly changes beyond a certain threshold, the proposed method starts sampling the P-V characteristic of the array in $60\% - 70\%$ of $V_{oc-mod}$(open circuit voltage of module) intervals, and at each sample, in case of sign change of $dP/dV$, P&O technique is utilized to determine the local peak. Finally, by comparing all peaks the GMPP is determined. The proposed method in [12] is also based on the method suggested in [11] and is similar to [13]. Dependency on $V_{oc-mod}$, a parameter that changes with environmental conditions, and low speed of the algorithm due to high sampling number are the weak points of these algorithms. The method proposed in [14] has good performance, but it is required to measure the voltage of each module. The method proposed in [15] is based on IC and sampling the P-V characteristic of the array in distances

M. A. Ghasemi, H. mohamadian and M. Parniani are with the Department of Electrical Engineering, Sharif University of Technology, Azadi Ave., Tehran, Iran (m_amingh@ee.sharif.edu, hosseinece@gmail.com, and parniani@sharif.edu).

of $0.8V_{oc-mod}$. It limits the search area for GMPP as in [11], and yields suitable results; but needs high sampling number. Reference [12] proposes two methods: the first approach samples the P-V curve and limits the search area based on short circuit current of the modules and the highest local power. As it is mentioned in [12], this method has high accuracy with low speed. Therefore, a second approach is proposed that estimates the local MPP power by measuring the currents of bypass diodes of the modules. Although the speed of tracking is improved, its implementation cost is high.

Studying MPPT as an optimization problem resulted in using evolutionary optimization methods such as particle swarm, simulated annealing and colony of flashing fireflies to find the GMPP [16-21]. In these methods, GMPP is obtained by sampling different points of the array P-V characteristic. These methods are mostly successful, but their sampling number is high. Since the GMPP can occur in a wide range of the P-V characteristic, initial sampling must cover the entire curve. Boost converters experience some transients to settle the voltage of the PV array [22, 23]. Then, as the sampling number increases, the speed of MPPT decreases. In [16, 19], a typical version of PSO algorithm is used that has low speed. In [18] the PSO method is modified to improve its speed and complexity. A method based on firefly algorithm is proposed in [20] that has better speed and efficiency in comparison to PSO-based algorithms. The proposed method in [21] uses the simulated annealing algorithm for MPPT under PSC. It is clear from the presented results that the samplings number is high and speed of GMPPT is even lower than the PSO-based method, while its accuracy is higher.

Generally, a good MPPT algorithm that is also successful in PSC should have the following properties:

1) Tracking the MPP rapidly for getting high efficiency,
2) Simple implementation with a low computational load,
3) Requiring less and cheaper sensors (removing current sensors of boost converter reduces the cost dramatically),
4) Imposing minimum disturbance to the connected grid.

Another issue that is less addressed in the literature is detection of PS occurrence. Before applying any MPPT process under PSC, it is necessary to detect its incidence. Until now, no special algorithm is presented to deal with this issue, and a sudden big change in the array power is commonly used as PS occurrence indicator [11, 12]. Determination of a threshold for big power change to distinguish between PSC and uniform irradiance condition (UIC) perfectly is not straightforward. Also, it is possible that in some situations, especially changing PS pattern, no big power change is observed. Another presented method is based on the fact that in PSC, there is big difference between the array currents in the low and high voltages of the array [18]. This method needs to sample from the array current in low and high voltages, and therefore imposes a big disturbance on the PV power and the connected grid.

In this paper, a novel MPPT algorithm is presented which is based on ramp change of the duty cycle and continuous sampling from the P-V characteristic of the array. Simple and cheap implementation due to its open loop operation, high and adjustable speed, robust and guaranteed performance in all conditions, and imposing minimum disturbance to the connected power system are advantages of the proposed method. Also, a new algorithm for detecting PSC occurrence on PV array is presented that has performance superiority over present methods.

The rest of the paper is organized as follows. The next section describes the characteristics of PV array in different conditions. In Sec. III, open loop control of the boost converter in the PV system is presented. The proposed algorithm for detection of PSC is described in Sec. IV. In Sec. V, the new MPPT method under PSCs is presented. Section VI verifies the proposed method with simulation and experimental results. Finally, the conclusions are made in the last section.

## II. Characteristics of PV Array

### *A. Uniform Irradiance Condition*

In the literature, different models are presented for solar cells. Among these models, single-diode model that is shown in Fig. 1 is used in this paper. Based on this model, relation between voltage (V) and current (I) of a PV module is expressed as follows:

$$I = I_{pv} - I_o\left[\exp\left(\frac{V + R_s I}{AV_T}\right) - 1\right] - \frac{V + R_s I}{R_{sh}} \quad (1)$$

where $I_{pv}$ is the equivalent photocurrent of module, $I_o$ is the reverse saturation current of the equivalent diode, A is the ideal factor, and $V_T\left(n_s\frac{kT}{q}\right)$ is the thermal voltage of module. Also, $R_s$ and $R_{sh}$ are the equivalent series and shunt resistances of the module. I-V characteristic of an array with $N_P$ parallel strings, each consisting of $N_s$ series modules, in UIC is then as follows.

$$I = N_p I_{pv} - N_p I_o\left[\exp\left(\frac{V + \frac{N_s}{N_p}R_s I}{AN_s V_t}\right) - 1\right] + \frac{V + \frac{N_s}{N_p}R_s I}{\frac{N_s}{N_p}R_{sh}} \quad (2)$$

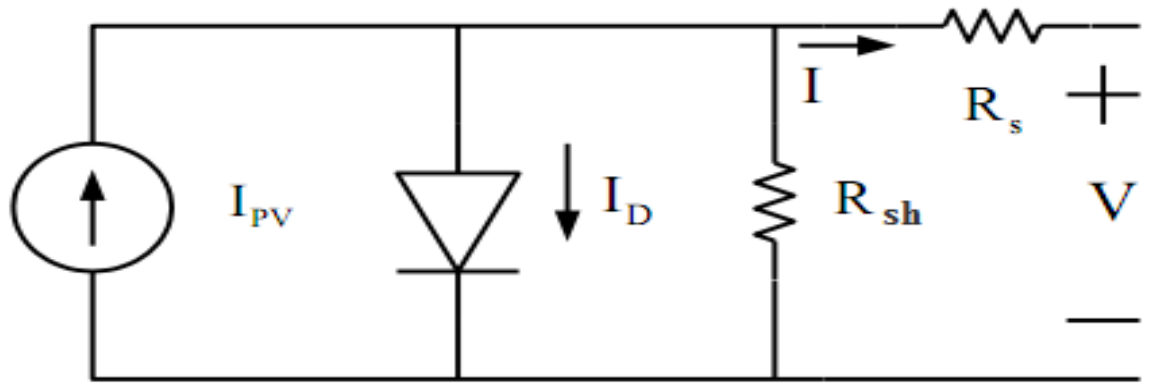

Fig. 1. Single-diode electrical model of a PV module.

In the rest of this paper the following symbol definitions are used. $V_{oc-mod}$ is open circuit voltage of PV module, $V_{oc-str}$ is open circuit voltage of PV string, $V_{oc-arr}$ is open circuit voltage of PV array, $V_{mpp}$ is the voltage of MPP, $V_{mpp-mod}$ is the voltage of module at its MPP, $V_{mpp-str}$ is the voltage of string at its MPP at UIC, and $V_{mpp-arr}$ is array voltage at MPP under UIC.

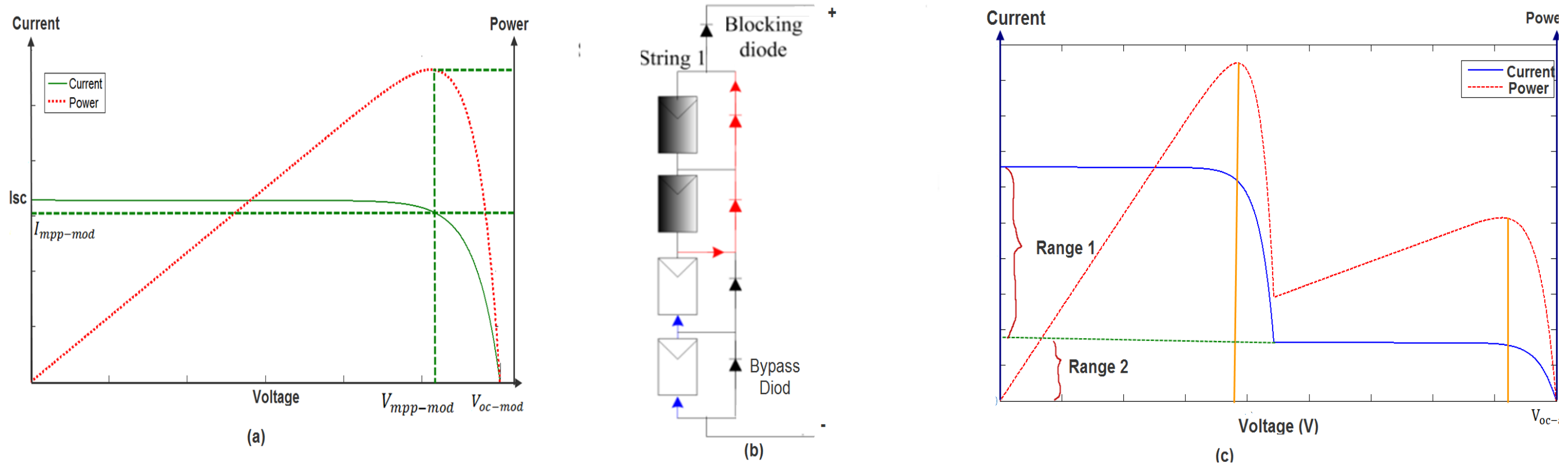

Fig. 2. (a) P-V and I-V characteristics of a typical PV module. (b) Structure of a sample shaded string. (c) P-V and I-V characteristics of the shaded string.

The module current is maximum at $V = 0$ and is known as short circuit current ($I_{sc}$). For voltages above $V_{oc-mod}$ there will be negative current, but a blocking diode will force it to zero. In Fig. 2 (a), I-V and P-V characteristics of a typical solar module under UIC are presented. In UIC, the maximum power point of module and array are unique and are achieved at $V_{mpp-mod} = \alpha V_{oc-mod}$ and $V_{mpp-arr} = \alpha V_{oc-arr} = N_s V_{mpp-mod}$, respectively; $\alpha$ is a coefficient that is dependent on model parameters of solar module.

### *B. Partially Shaded Condition*

For simplicity, it is initially supposed that the array under PSC is subjected to two different irradiance levels. Modules that receive high irradiance level ($HS$) are called insolated modules and those which receive lower irradiance level ($LS$) are named shaded modules.

The insolated modules of a string drive the string current. Therefore, portion of the string current that is greater than the generated current of shaded modules passes through parallel resistance of the shaded modules and generates negative voltage across them. Thus, the shaded modules consume power instead of generating it. In this condition, not only the overall efficiency drops, but also the shaded modules may be damaged due to hot spots. To prevent this condition, a bypass diode is connected in parallel to each module, to let the extra current of the string pass through it. Consequently, the voltage across that module will be about $-0.7\,V$ and efficiency of the string will improve. The structure of a sample shaded array is shown in Fig. 2 (b). Further details about the modeling of array in PSC are given in [3, 24].

### *C. Critical Observations under Partially Shading Condition*

0Figure 2 (b) and (c) show the structure and I-V and P-V characteristics of a typical partially shaded string with $N_s = 4$ series modules, $n_{sh} = 2$ shaded modules, and $n_{in} = 2$ insolated modules. As explained in the previous subsection, for currents higher than $I_{sc}$ of shaded modules (Range 1), their bypass diodes conduct extra current and cause the voltage across them to be about $-0.7$ to $-1\,V$. In this situation, the string voltage is equally divided only between the insolated modules. For currents lower than $I_{sc}$ of the shaded modules (Range 2), insolated modules operate in approximately constant voltage area, and therefore, the voltage across each of these modules will be more than $V_{mpp-mod}$ and close to $V_{oc-mod}$. The P-V characteristic of the string has two MPPs. The first one is at $V_{mpp-1} \approx n_{in} V_{mpp-mod} - n_{sh} * 0.7$ and the second MPP occurs when the voltage of one shaded module is about $V_{mpp-mod}$. The string voltage in this local MPP ($V_{mpp-2}$) is bound as follows:

$$N_s\, V_{mpp-mod} < V_{mpp-2} < n_{sh} V_{mpp-mod} + n_{in}\, V_{oc-mod} \qquad 3)$$

When the irradiance ratio $IR = HS/LS$ decreases, $V_{mpp-2}$ gets close to the lower bound of (3), and as it increases, $V_{mpp-2}$ moves toward the upper limit. Also, when $K = n_{sh}/n_{in}$ is too high the upper limit of (3) approaches the lower limit, and $V_{mpp-2}$ gets close to it. According to the above discussion, it can be shown that in one string, the minimum difference between the voltages of two local MPPs is more than $V_{mpp-mod}$.

## III. Open Loop Control of Boost Converter in PV System

In Fig. 3, a two-stage grid connected solar system is shown. In the first stage, DC/DC boost converter plays the main role in absorbing power from the PV array by controlling its voltage. In the second stage, an inverter controls the output voltage of the DC/DC converter and generates AC voltage to connect the solar system to the grid. Because of the DC link capacitor between the boost converter and the inverter, there is little coupling between the two stages and the stages can be studied separately [25].

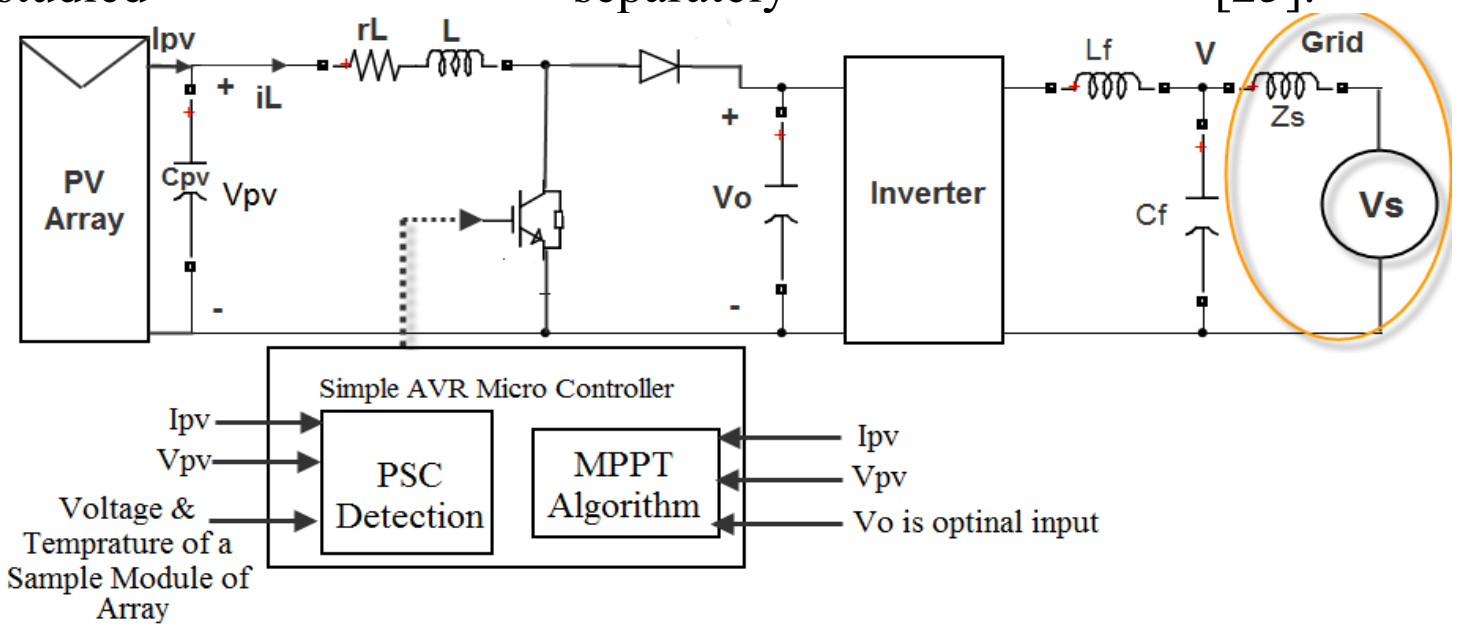

Fig. 3. Overview of a two-stage grid connected PV system structure.

Generally, there are two control approaches for regulation of a PV array using boost converter; i.e. close loop and open loop controls. Reference [23] shows that in a PV array connected to the boost converter, the worst case from stability and dynamic response points of view, occurs when the array operates in constant current region and low irradiance level, where dynamic resistance of the array has its largest negative value. Due to dependency of the system dynamic response to the operating point and environmental conditions, it is not possible to control the array voltage in close loop fashion using a single-loop PI voltage controller properly, and another inner control loop is required (boost converter inductor current loop) to reach desired dynamic response of the system (high speed, low transient, and zero steady state error) [23]. This two-loop control method needs two PI controllers and an expensive current sensor. In contrast, in open loop control, which is a common method for boost converters control, there is no feedback, and the appropriate input voltage is generated considering the relation between the input voltage ($v_{in}$) and output voltage ($v_o$) of the converter as in (4).

$$v_{in} = v_{pv} = (1-D)v_o \quad (4)$$

In this method, it is not necessary to measure the boost converter inductor current and an expensive current sensor is saved. However, the system response may have some steady state error and more transients than the close loop method. One of the important parameters in MPPT of a PV system is the sampling time. After applying a new command voltage ($v_{in}^{ref}$) to the converter, to prevent instability and disruption in MPPT, sampling from the array voltage and current must be done after settling the system transient response. Therefore, sampling time period must be more than this settling time.

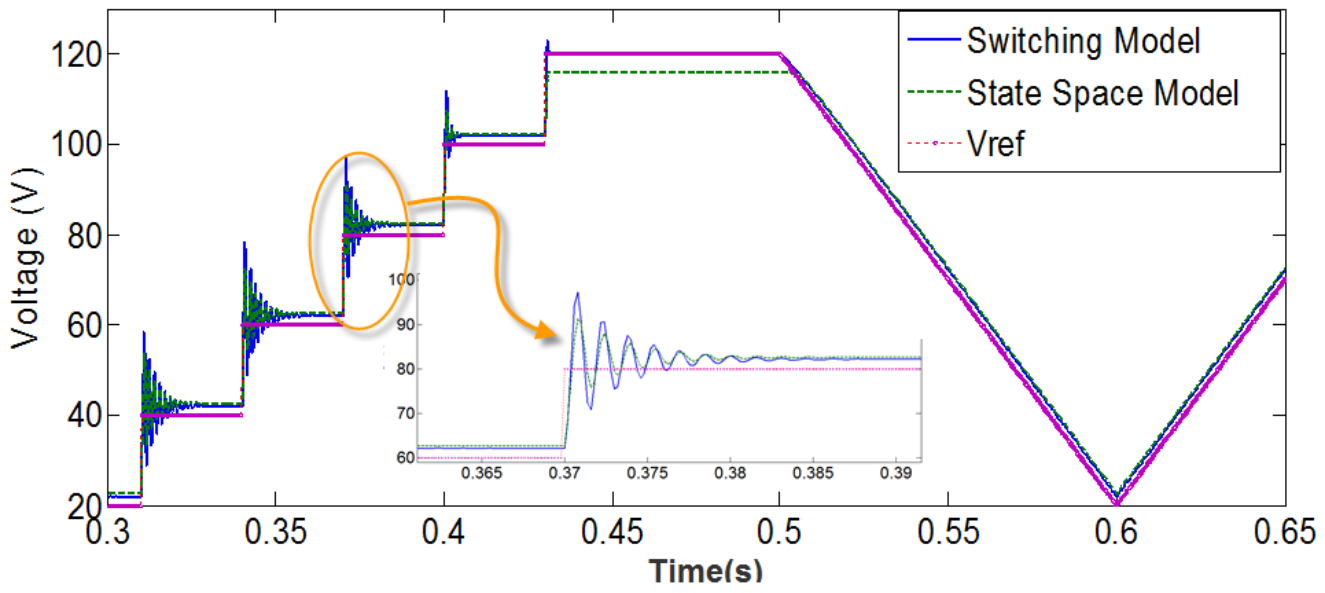

Fig. 4. Response of switching and averaged state space models of boost converter in PV system to step and ramp commands.

For further analysis, response of a PV array connected to a boost converter with open loop control is studied through simulation in Matlab/Simulink environment. Converter parameters are presented in Table I and the simulated PV array has $V_{oc-arr} = 130\,V$ and $I_{sc} = 8\,A$. Output voltage of the boost converter is also considered constant at 250V.

TABLE I. BOOST CONVERTER PARAMETERS

| $r_L(\Omega)$ | $L(uH)$ | $C_{pv}(uF)$ | $Switching\ Frequency\ (kHZ)$ |
|---|---|---|---|
| 0.3 | 600 | 100 | 20 |

Both the switching and averaged state space models of the system are simulated and their responses to step and ramp command signals by open loop control are shown in Fig. 4. Following conclusions are made from the system response:

1) Responses of the accurate switching model and the averaged state space model are almost identical.

2) The system response to step and ramp command signals contain some steady state error. This error can deteriorate the MPPT methods that are based on sampling from specific points of the array's P-V characteristic [13].

3) Oscillation, overshoot and settling time of the system to step commands is high, especially when the operating point in in the constant current region of the PV array, which impose higher switching stress and losses. In contrast, the ramp response has negligible transient.

4) Settling time of the system step response is about 15ms. Thus, for MPPT application, sampling time must be more than 15ms. It is noteworthy that $r_L$ is considered high, while in practice, for better efficiency, it is lower and results in higher settling time.

## IV. PARTIAL SHADING CONDITION DETECTION

In this section, an algorithm for PSC detection is presented which is based on three criteria. Also, performance of the final algorithm is evaluated in various PS patterns.

### *A. PSI index as Partial Shading Condition Detection Criterion*

The first proposed criterion is based on a new index that is defined as follows:

$$PSI = \frac{\Delta P}{\Delta V.P}|_{V_{mpp-arr}} = \frac{\frac{\partial P}{\partial V}}{V.I}|_{V_{mpp-arr}} = \frac{\frac{\partial (VI)}{\partial V}}{V.I}|_{V_{mpp-arr}} = \frac{1}{V_{mpp-arr}} + \frac{\partial I}{I.\partial V}|_{V_{mpp-arr}} \quad (5)$$

The criterion is normalized derivative of the PV array power respect to the array voltage at $V_{mpp-arr} = N_s V_{mpp-mod} = V_{mpp-str}$, which is similar to that used in IC method for MPPT. At UIC, PSI is zero. Under PSC, however, the local MPP voltage changes from $V_{mpp-arr}$, and therefore, PSI is not zero and is dependent on the shading pattern.

According to Sec. II, when a PV string is under PSC, the voltage across the shaded module ($V_{mod-shaded}$) at $V_{mpp-str} = N_s V_{mpp-mod}$ is bound as follows:

$$\frac{N_s V_{mpp-mod} - n_{in} V_{oc-mod}}{N_s - n_{in}} < V_{mod-shaded} < V_{mpp-mod} \quad (6)$$

From (6) two cases may arise for $\frac{\partial I}{I\,\partial V}|_{V_{mpp-arr}}$:

1) $\boldsymbol{N_s V_{mpp-mod}} > n_{in} \boldsymbol{V_{oc-mod}}$ (Fig. 5. Case1):

In this condition, $V_{mod-shaded}$ is positive and the absolute value of $\frac{\partial I}{I\,\partial V}|_{V_{mpp-arr}}$ is less than its value in UIC, the local MPP of the string is in $V > V_{mpp-arr}$, and PSI is positive.

2) $\boldsymbol{N_s V_{mpp-mod}} < n_{in} \boldsymbol{V_{oc-mod}}$ (Fig. 5. Case2):

In this case the shaded modules are bypassed with the bypass diodes, and $V_{mod-shaded} \approx -(0.7{\sim}1)V$. The insolated modules operate in the constant voltage region. Therefore, $\frac{\partial I}{I\,\partial V}|_{V_{mpp-arr}}$ is much bigger than its value in UIC; PSI is negative and local MPP of the string is in $V < V_{mpp-arr}$.

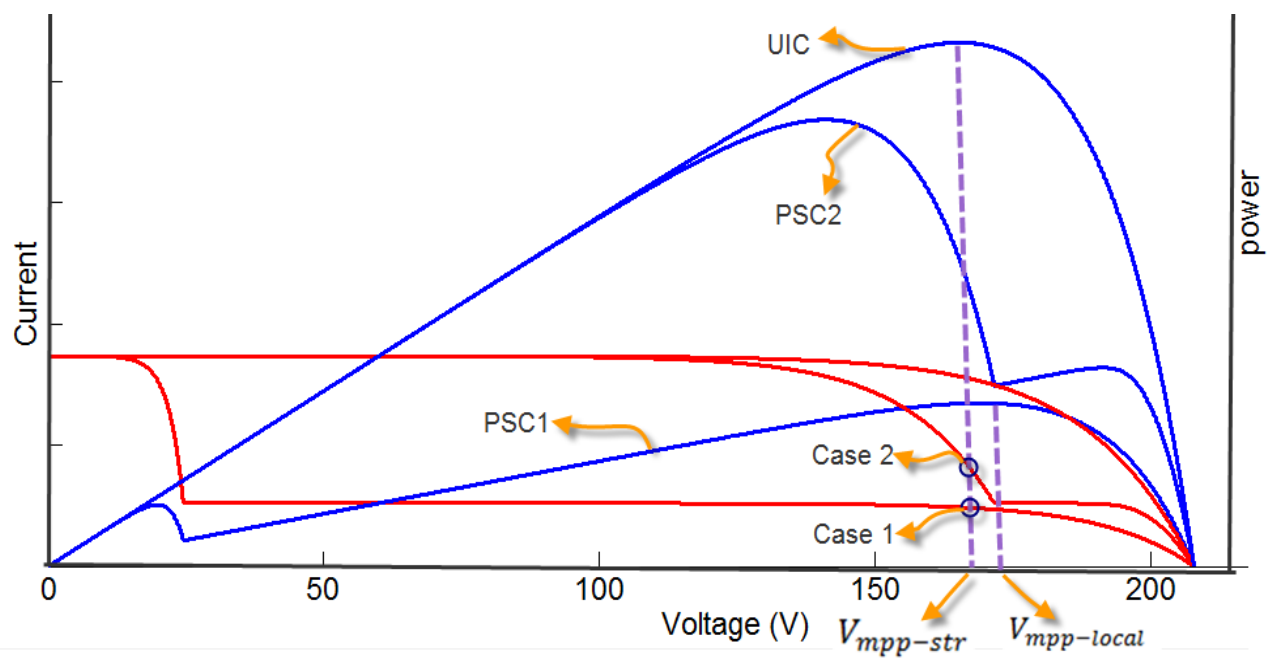

Fig. 5. I-V and P-V characteristics of PV string in different PSCs.

To investigate the effectiveness of the PSI index in PSC detection, behavior of a sample string, as a representative of an array, is analyzed in different PS patterns. For simplicity and without loss of generally, only two irradiance levels are considered in PSs.

According to the results of Sec. II and (3), it can be easily shown that in a shaded string, when $\boldsymbol{K = n_{sh}/n_{in}}$ is too high, the second local MPP ($V_{mpp-2}$) will be near $V_{mpp-str}$. Therefore, the PSI index may be near zero and PS detection fails. The same result may be yielded when $\boldsymbol{IR = HS/LS}$ is too low. Although the proposed algorithm may rarely mistake in detection of PSCs in the above-mentioned situations, but the main objective, which is GMPPT, is not lost. To prove this fact, a sample string under PSC (such as the PSCs in Fig. 5) is considered that has two local MPPs; the first one is in the range $V_{mpp-1} < V_{mpp-str}$ and has $P_{mpp-1} \approx n_{ni}\, V_{mpp-mod}\, I_{mpp-1}$, and the next one is in $V_{mpp-2} > V_{mpp-str}$ with the following power relation:

$$P_{mpp-2} = V_{mpp-2} I_{mpp-2} \approx V_{mpp-2}\cdot\left(\frac{1}{IR}\right).I_{mpp-1} > (K+1)\, n_{in}\, V_{mpp-mod}\cdot\left(\frac{1}{IR}\right).I_{mpp-1} = (K+1)\left(\frac{1}{IR}\right)P_{mpp-1} \quad (7)$$

It is obvious that when $IR$ is too low or $K$ is too high (the same situation that PSI index may be near zero, e.g. PSC1 in Fig. 5), $P_{mpp-2}$ will be much greater than $P_{mpp-1}$. Therefore, if the PSI index mistakes in detection of this PSC, the conventional P&O algorithm used in the UICs tracks the second MPP which is the GMPPT.

### *B. Updating $V_{mpp-arr}$ and Final PS Detection Criteria*

Until now, it was supposed that $V_{mpp-arr}$ is available for PSI evaluation. In practice, $V_{mpp-arr}$ and $V_{mpp-mod}$ are dependent on the type of modules and temperature as in (8); and also, there is some difference between the temperatures of the shaded and insolated modules.

$$V_{mpp-arr} = V_{mpp-arr-SC} * \left(1-\rho_{arr}.(T-25)\right) = \sum_{i=1}^{N_s} V_{mpp-mod-SC} * (1-\rho_{mod-i}.(T_i-25)) \quad (8)$$

where $V_{mpp-arr-SC}$ and $V_{mpp-mod-SC}$ are $V_{mpp-arr}$ and $V_{mpp-mod}$ in standard condition (S=1$kW/m^2, T = 25\ C^{\cdot}$), respectively. T is temperature and $\rho_{arr}$ and $\rho_{mod}$ are the temperature dependency coefficients of $V_{mpp-arr}$ and $V_{mpp-mod}$, respectively. In the UIC, the operating voltage of the array is $V_{mmp-arr}$. Therefore, $V_{mmp-arr}$ is available continuously. Also, its slight dependence on irradiance level can be updated easily, using the array current at $V_{mmp-arr}$. Under PSC, the operating voltage is not $V_{mmp-arr}$. Consequently, $V_{mmp-arr}$ that is dominantly dependent on the temperature of the array is not available.

If the PS is due to relatively fast transient phenomena like the passing clouds, the temperature cannot change rapidly, and therefore, it is almost identical in all modules. Otherwise, temperatures of the shaded and insolated modules are different; and this temperature difference is proportional to the difference of the radiation levels. Hence, for updating $V_{mmp-arr}$, temperatures of all modules must be measured, a requirement that is not economical. Hence, in the proposed algorithm only the temperature of one sample module is used for updating $V_{mmp-arr}$ according to (8) ($V_{mmp-arr}$ and $V_{mmp-mod}$ are updated using $\rho_{arr}$ and $\rho_{mod}$, respectively). In this situation, it is not clear whether the sample module is insolated or shaded. Accordingly, three cases may be fronted as follows:

1. The whole array is in UIC, and therefore, the temperature of all modules is the same as the sample module temperature. Thus, there is no error in updating $V_{mpp-arr}$ in this case, and UIC can be detected using the PSI index.
2. The array is in PSC and the sample module is insolated. In this case because of the negative value of $\rho_{mod}$ for all types of modules and $\rho_{arr}$ in (8), the updated value of $V_{mpp-arr}$ will be less than its real value. Therefore, the calculated value of PSI and the difference between the real local MPP voltage and the updated $V_{mpp-arr}$ (named $|\Delta V_{mpp-arr}|$) will be greater than its real value. Hence, PS detection becomes easier.
3. The array is in PSC and the sample module is shaded. In this case, also because of the negative value of $\rho_{arr}$, the updated value of $V_{mpp-arr}$ is greater than its real value. Therefore, the calculated value of PSI and the difference between the local MPP voltage and the updated $V_{mpp-arr}$ will be smaller than its real value and may be even zero. Therefore, success of the proposed algorithm may be affected. In this situation, voltage of the sample module is measured while the array voltage is at updated value of $V_{mpp-arr}$. Clearly, voltage of the sample module at this point is quite different from the updated value of $V_{mpp-mod}$ (named $|\Delta V_{mpp-mod}|$) when the array is under PSC. Otherwise, their difference will be nearly zero. This modification ensures success of the proposed algorithm for PS detection.

According to the above discussion, the PSI index criterion is reinforced with two other criteria. These two criteria are defined based on normalized values of $|\Delta V_{mpp-arr}|$ and $|\Delta V_{mpp-mod}|$ that are defined in the above. Finally, the criteria for PS detection will be as follows.

$$|PSI| > 0.001$$
$$\left|\frac{\Delta V_{mpp-arr}}{V_{mpp-arr}}\right| > 0.02 \qquad (9)$$
$$\left|\frac{\Delta V_{mpp-mod}}{V_{mpp-mod}}\right| > 0.02$$

The specified thresholds in (9) are determined according to the simulations of many PS scenarios on various structures of PV array. Based on these criteria, the array is in the PSC if at least one of these conditions is met. In Fig. 6, flow chart of the proposed algorithm for PS detection is shown. The proposed PSC detection does not impose any considerable disturbance on the system, since PSI is evaluated at $V_{mpp-arr}$.

It is noteworthy that $\rho_{arr}$ and $\rho_{mod}$ may be non-identical because the module models in the array may differ. Also, $\rho_{arr}$, $\rho_{mod}$, $V_{mpp-arr-sc}$ and $V_{mpp-mod-sc}$ may change due to aging. Nevertheless, they can easily be updated online when the array is under UIC. For the sake of brevity, their updating process is not explained here. However, it can easily be shown that the effectiveness of the algorithm is independent from uniformity of modules and their aging.

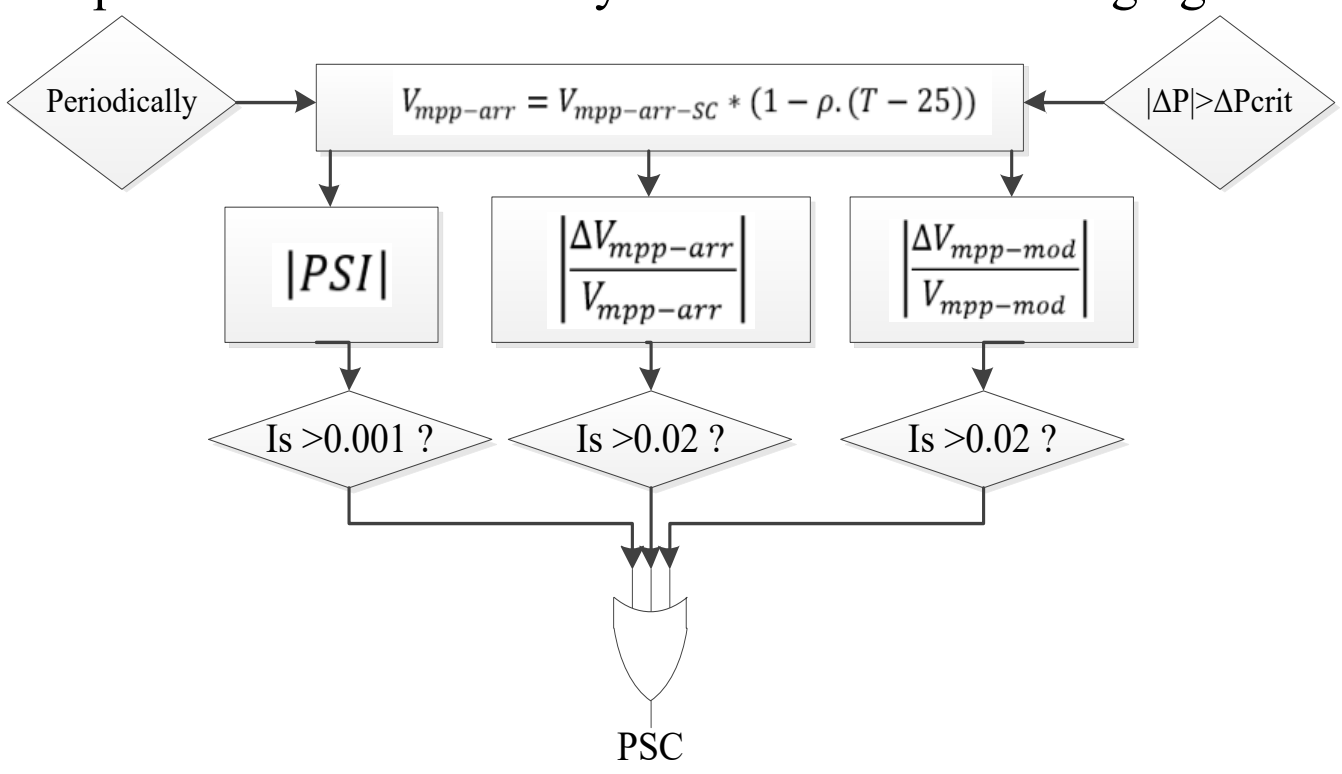

Fig. 6. Flow chart of the proposed algorithms for PS detection.

So far, the proposed algorithm is studied in a string of series modules. In (10), it is shown that PSI of an array is the weighted average PSI of individual strings, and therefore, using the PSI and two other criteria in (9) suffices for PS detection in any array.

$$PSI = \frac{\Delta P}{\Delta V.P}|_{V_{mpp-array}} = \frac{\sum \Delta P_i}{\Delta V.\sum P_i}|_{V_{mpp-array}} = \sum PSI_i \frac{P_i}{\sum P_i} \qquad (10)$$

where $P_i$ and $\Delta P_i$ are the power of string $i$ and its differentiate, respectively.

### *C. Effectiveness of proposed algorithm for PSC detection*

In the following, a sample array is simulated in different PSCs to verify the effectiveness of the proposed PS detection algorithm.

TABLE II. ELECTRICAL DATA OF MODULE ND195R1S IN STANDARD TEST CONDITION

| $P_{max}$ | $V_{oc}$ | $I_{sc}$ | $V_{mpp}$ | $I_{mpp}$ | $P_{max}$ $Termal\ Coefficeint$ | $\rho_{mod}$ |
|---|---|---|---|---|---|---|
| 195 | 29.7 | 8.68 | 23.6 | 8.27 | -0.44 $\%/C^{\cdot}$ | -0.329$\%/C^{\cdot}$ |

As shown in 0, the array is 3x5, composed of ND195R1S modules with electrical data given in Table II. In this simulation, the modules are under three different irradiations with the following associated temperatures: ($S_1 = 0.9\ kw/m^3$,$T_1 = 35C^{\cdot}$), ($S_2 = 0.6\ kw/m^3, T_2 = 30C^{\cdot}$) and ($S_3 = 0.3\ kw/m^3, T_3 = 25C^{\cdot}$). Series and shunt resistances of the modules are also considered in the simulations.

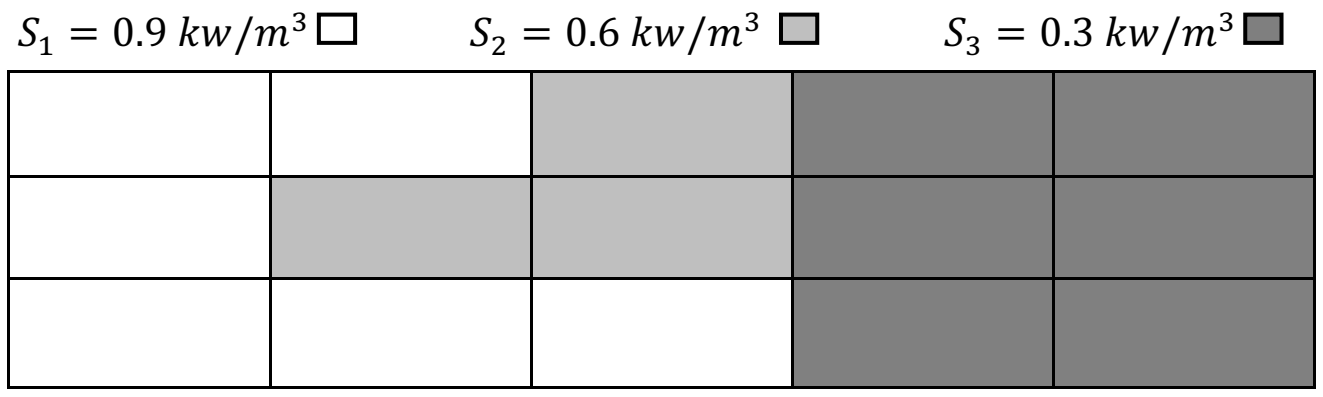

Fig. 7. The simulated array configuration (indicates the sample module).

Results of five different PSC simulations are presented in Table III. Patterns of the PSCs on the array is presented with a ternary digit for each string of the array. These digits are the number of the modules in each string with ($S_1, T_1$), ($S_2, T_2$) and ($S_3, T_3$), respectively. In these simulations, it is assumed that temperature and voltage of the marked module in 0 is measured. As it is clear from Table III, the PSI index fails only in detection of PSC5, and the third criterion fails only in detection of PSC2. But using all criteria in (9) makes the algorithm successful in all cases. The proposed method has complete success in detection of all simulated PS patterns. Considering the three criteria in (9) has resulted in good robustness of the method. As it was mentioned previously, robustness of the proposed method is reduced only under the PSCs that $K$ is too high and $IR$ is too low, and it may be possible that the proposed method does not detect the PSC. In these situations, as proved in part B of this section, the local MPP, which is near $V_{mpp-arr}$, is the GMPP and the conventional P&O algorithm tracks it. Therefore, the final goal that is GMPPT is not missed.

TABLE III. RESULTS OF PSC DETECTION IN SAMPLE PSCS

| | PSC1 | PSC2 | PSC3 | PSC4 | PSC5 |
|---|---|---|---|---|---|
| PS Patterns | 2-2-1 | 5-0-0 | 0-1-4 | 1-1-3 | 1-1-3 |
| | 1-3-1 | 3-1-1 | 0-0-5 | 1-1-3 | 5-0-0 |
| | 3-2-0 | 3-2-0 | 1-1-3 | 1-0-4 | 4-0-1 |
| $\|PSI\|$ | 0.008 | 0.0036 | 0.002 | 0.003 | 4.00E-04 |
| $\left\|\frac{\Delta V_{mpp-arr}}{V_{mpp-arr}}\right\|$ | 0.09 | 0.04 | 0.022 | 0.03 | 0.003 |
| $\left\|\frac{\Delta V_{mpp-mod}}{V_{mpp-mod}}\right\|$ | 0.3 | 0 | 0.03 | 0.08 | -0.08 |
| PSC | yes | yes | yes | yes | yes |

It is worth comparing the proposed PSC detection method with that of [11, 12]. Their method is based on observing a big power change, and is sensitive to the relevant threshold: a smaller threshold may cause wrong detection of PSC, and a bigger one may result in missing it. In contrast, the proposed method in this paper is activated in two ways: 1) periodically, 2) after observing a noticeable power change. For perfect detection of PS, the threshold of this power change can be set to lower values, because after observing the change, the criteria in (9) will be examined to rule out wrong PSC candidates. Also, in comparison to the method used in [18], which samples the array current in low and up voltages to detect PSC, the proposed method in this paper does not impose any big disturbance on the system as the method of [18] does.

## V. Proposed Algorithm for MPPT Under PSC

Heuristic algorithm based methods such as PSO, as well as most of other methods for MPPT in PSC, need to sample the P-V characteristic of the array in different voltages of the search area. Noting to the settling time of boost converter to step commands, these methods have low speed in GMPPT.

MPPT is a time varying optimization problem, in which the objective function evaluation is done physically; i.e. by applying specific voltages to the array, its output power is measured after settling its voltage, whereas in the numerical optimization problems, function evaluation is done numerically and imposes calculation burden on the processor. As mentioned in Sec. III, sampling time period for MPPT must be greater than the settling time of the boost converter. This settling time depends on the design and operating point of PV array. Maximum settling time of the boost converter used in experiment and simulations of this paper is about 20ms.

According to Sec. II, under PSC, the GMPP is in the following voltage region that must be searched for GMPPT:

$$V_{mpp-mod} < V < V_{oc-arr} \quad (11)$$

A straight solution for GMPPT with minimum steps is that sampling from P-V characteristic of the array be done only in specific points [13]. In practice, these methods rely on approximations and cannot guarantee the GMPPT.

When sampling from the array power is done in $V_1$ and $V_2$, respectively; indeed the array voltage experiences all voltages between $V_1$ and $V_2$ continuously. This is due to the fact that the voltage of the parallel capacitor with the array cannot change steeply. Therefore, almost in all of the MPPT methods, the array experiences all voltages of (11).

According to the above discussion, two important facts inspire using ramp voltage as the command signal of converter to search the voltage region of (11) for GMPPT:

1. In contrast to the response of the boost converter to step commands, settling time and transient of the boost converter to ramp command is nearly zero (Fig. 4).
2. PV arrays do not have considerable dynamics and can be assumed static. Unlike dynamic systems, then, the measured power at each moment is related to the array voltage at the same moment, corresponding to a point on the P-V characteristic of the array.

Thus, the concept of scanning I-V characteristic of the array with adjustable high speed ramp command voltage (or ramp change of duty cycle) is proposed in this paper. Along with this ramp input, the array voltage and current is sampled continuously with proper rate.

In two different situations that may occur for the PV array, the proposed algorithm for GMPPT operates as follows:

a) While the array is under UIC and operates at ($V_1 = V_{mpp-arr}, P_1$), a PS occurs and the operating point changes to ($V_2 = V_1, P_2$). The proposed PSC detection algorithm is initiated by this power change, and determines whether the array is still at UIC or has undergone PSC. If no PS is detected, $P\&O$ algorithm is called. Otherwise, the proposed MPPT algorithm is activated. Based on the proposed MPPT method, the maximum array power during MPPT process and its corresponding voltage ($V_e, P_e$) are initialized with ($V_2, P_2$). Then, a positive ramp voltage command, starting from $V_1$, is applied to the boost converter according to (4). For this purpose, the duty cycle can also be changed with ramp function, without need to know the output voltage ($v_o$) of the boost converter. Consequently, the array voltage changes ramp-likely as shown in Fig.10 (a). Simultaneously, the array voltage $V(t)$, which may have some error from the command voltage, and its power $P(t)$ are sampled as ($V_s = V(t), P_s = P(t)$). At each moment, if $P_s$ is greater than $P_e$, then ($V_e, P_e$) is updated with ($V_e = V_s, P_e = P_s$). This process continues until the array voltage reaches to $V_{oc-arr}$. Then, the command voltage ramp sign is inverted and the array voltage is reduced ramp-likely. Updating ($V_e, P_e$) is continued until the voltage reaches to $V_{mpp-mod}$. Finally, GMPP of the array will be the final value of the ($V_e, P_e$). Then, the array voltage is drived to $V_e$ and the P&O algorithm is called to resume the local MPPT around this operating point.

b) The array is under PSC and shading pattern changes. In this case, based on the proposed concept, ramp voltage command is applied to the converter to bring the array voltage to $V_{mpp-arr}$. At this point, PS detection criteria is checked to determine if the array is at UIC or under PSC. If no PS is detected, P&O algorithm is called. Otherwise, MPPT process is started by applying positive ramp voltage command to the converter. The rest of the process is same as explained in (a).

To limit the search region for GMPPT, further analyses are presented as follows.

1- Assume a sample operating point of the array as ($V_s, I_s$). It is known that when the array voltage increases, its current decreases. Therefore, the array current ($I_{arr}$) for $V > V_s$ is lower than $I_s$. Hence,

$$P_{arr}(V > V_s) = VI_{arr} < VI_s \quad (12)$$

In addition, because the maximum voltage of the array is $V_{oc-arr}$,

$$P_{arr}(V > V_s) < V_{oc-arr}I_s \quad (13)$$

Based on the above arguments, during positive ramp command, at any point in which $V_{oc-arr}I_s$ is less than the last updated value for MPP ($P_e$), the array power will also be less than $P_e$ at all upper voltages. Therefore, continuing the positive ramp command is not required. In other words, the search region will be limited to $V_s$ in which,

$$V_{oc-arr}I_s < P_e \quad (14)$$

2- Whenever PS occurs after a UIC, negative ramp of MPPT process is bound as follows. For the voltages that $VI_{sc} < P_e$ the array power is less than $P_e$ and it is not needed to search this region. Hence, lower voltage of search region will be $V_s$ that,

$$V_s < P_e/I_{sc} \tag{15}$$

Besides, MPP current of PV arrays under UIC is about $0.9I_{sc}$[3], and therefore $I_{sc}$ is approximately known in term of $I_{mpp-arr}$.

One notes that the proposed MPPT method guarantees convergence to the GMPP under any partial shading condition. The reason is that the voltage region (11) is considered completely, and sampling from the voltage and power of the array is done in the entire region, not at some special points. Furthermore, the proposed MPPT method does not need any electrical characteristics of the PV array except to $V_{oc-arr}$ which is used to define the search region. All MPPT methods needs to know $V_{oc-arr}$ to know the search region. Besides, the exact value of $V_{oc-arr}$ is not necessary, and its approximate value can be determined in term of $V_{mpp-arr}$.

Flow chart of the proposed algorithm for MPPT in PSC is shown in Fig. 8. The ramp voltage can be implemented either with an analog rate limiter or digitally with small step changes in duty cycle.

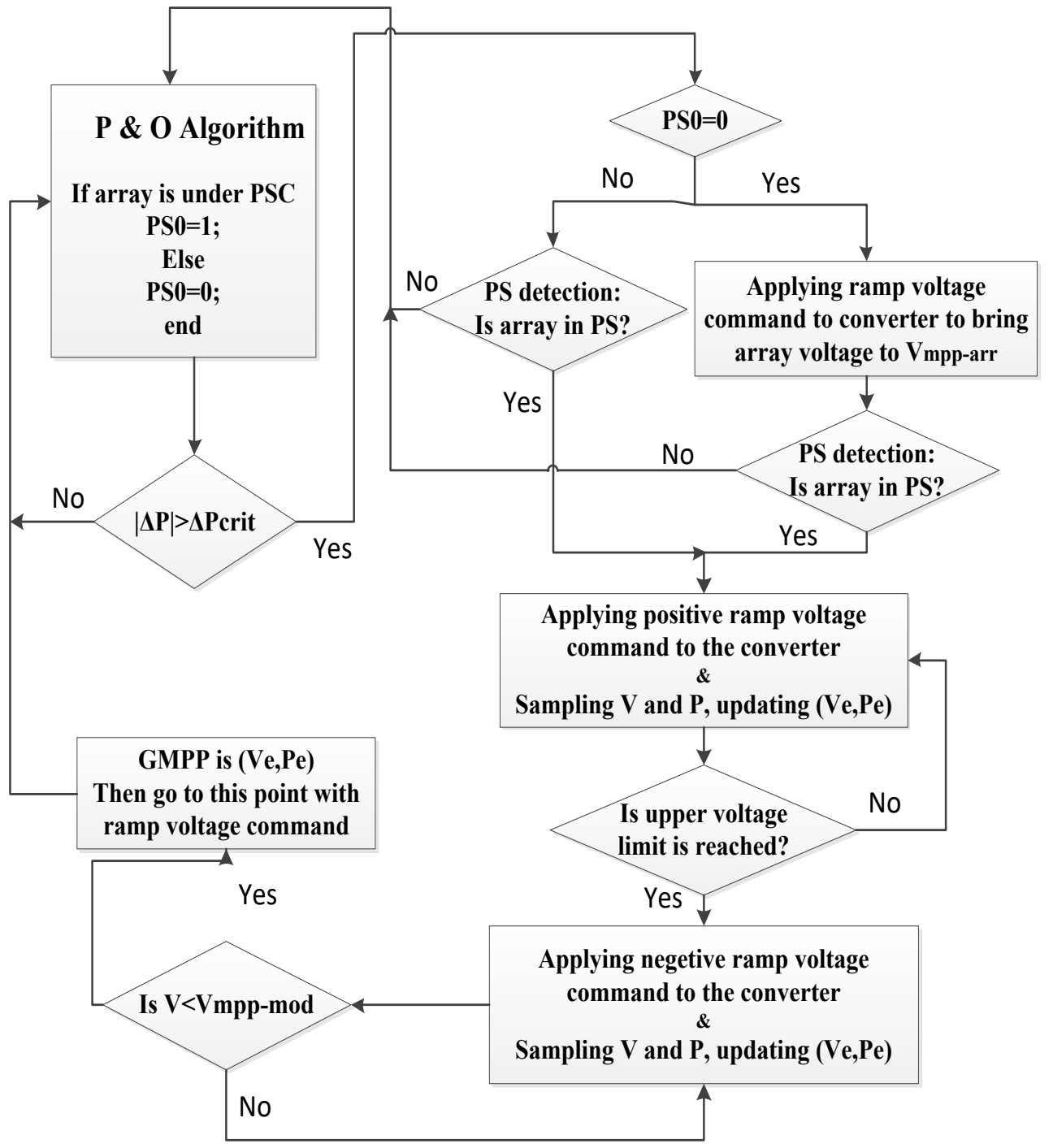

Fig. 8. Flow chart of the proposed algorithms for MPP tracking under PSC.

For selection of the ramp rate, the following points are noted:

1- Sampling rate of the array voltage and current must be coordinated with the ramp rate of command voltage. For example, suppose a PV array with $V_{oc-arr} = 200V$. To reach the GMPP voltage at about 50ms, the voltage ramp is selected $R = 200/0.05 = 4000\,\mathrm{V/s}$. To achieve the GMPP voltage with maximum 1V error, at least one sample per each 2V interval is needed. Therefore, the minimum sampling rate must be

$$f_{sam} > \frac{4000}{200} * \frac{200}{2*1} = 2kHz \tag{16}$$

Since the existing micro-controllers have much faster sampling capability, no serious limitation on voltage ramp is imposed from this aspect.

2- PV arrays have very fast dynamics because of their current leakage to ground, which are negligible and ignored in MPPT process. Theoretically, using very excessive ramp rates in the range of those fast dynamics, may deteriorate performance of the proposed method. However, this is impractical, and as mentioned earlier, the PV arrays are dealt as static systems.

3- When a ramp command voltage is applied to the boost converter, its voltage changes in proportion to the ramp rate of command voltage, but it may have some steady state error and very little transients (Figs. 4, 10, 11). However, in the proposed algorithm, sampling from the array voltage and current is done continuously, and it is not necessary that the array voltage be the same as the command voltage. Therefore, imperfect response of the boost converter does not imply any limitation on the proposed method.

The last concern is that excessive increasing of the ramp rate results in high $dP_{pv}/dt$ during GMPPT, which can disturb the connected grid. Nevertheless, as it is shown in the next section, selecting the voltage ramp as $4000\,V/s$ results in fast MPP tracking and sufficiently low disturbance.

## VI. Simulation and Experimental Results

In this section, performance of the proposed method for GMPPT in PSC is evaluated in various aspects using simulations and experiments.

### *A. Simulation Results*

The proposed method is compared with the PSO-based algorithm with 3 primary particles presented in [19] using simulation in Matlab/Simulink software. It is also compared with the method of [11], which is one of the best intelligent MPPT methods that is referenced very often. The simulated system configuration was described in Sec. III. All parts of the presented system in Fig. 3 are considered in the simulations. For the sake of brevity, control of grid-connected inverter is not described here. The PV array is a 5x6 array composed of ND195R1S modules with electrical data given in Table II. P-V & I-V characteristics of the simulated array under the UI and two PS patterns are shown in Fig. 9. In this study, the array is initially under uniform radiance, and then it passes to the PSC patterns.

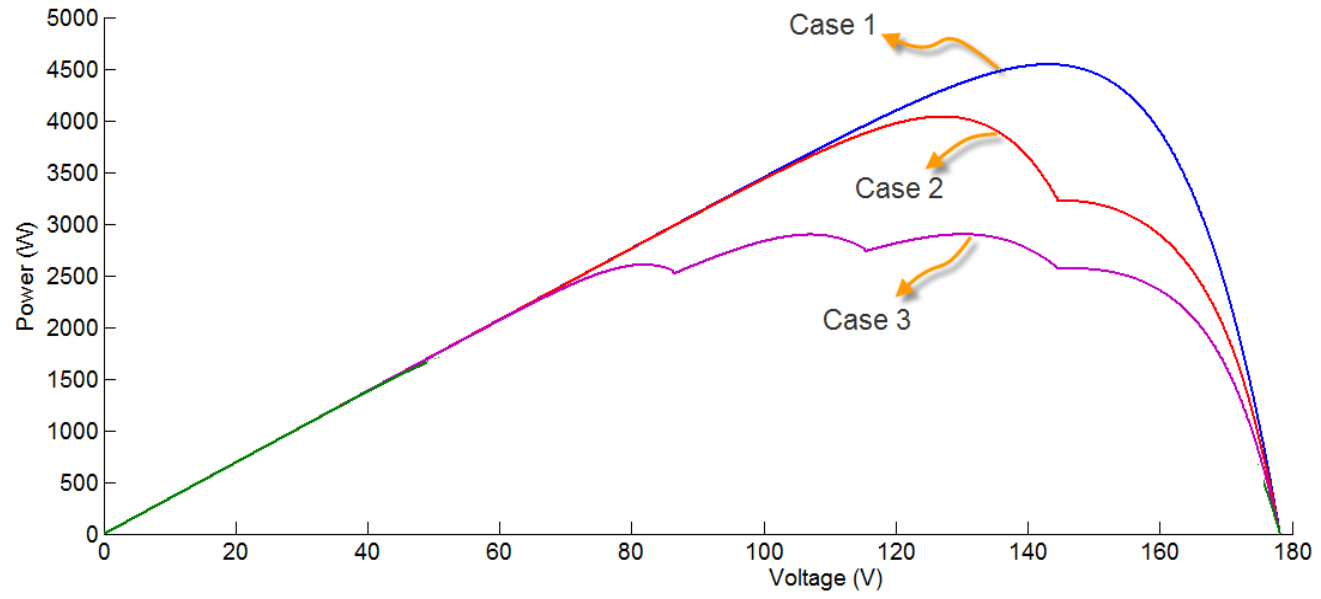

Fig. 9. P-V characteristics of array in UIC and two different PSCs.

Simulation results are presented in Fig. 10 (a-d), and the efficiency of the proposed MPPT algorithm is compared with the PSO-based method in [19] and the proposed method in [11]. In the simulation of the proposed method, analog to digital conversion time of processor in measuring the voltage and current of the array is considered 0.5ms, which is achievable by a low-speed micro controller. Also, the voltage ramp for searching the GMPP is set to 4000 V/s. The proposed algorithm tracks the GMPP in all cases rapidly in lower than 70ms. Because of the lower speed of two other methods, the time interval between PSC pattern changes is increased to 2s for them.

Changing the array voltage with ramp, not only increases the speed of tracking and reduces its transients and stress on the converter, but also has extra benefits in terms of interaction with the connected grid. In the grid-connected PV system (Fig. 3), the inverter must deliver all generated PV power to the grid rapidly for regulating the voltage of the output capacitor of boost converter ($v_o$). Therefore, changing the array power leads to changing the injected power to the grid, and it yields voltage transients at the point of grid connection ($V$ in Fig. 3). Step changes in the array voltage, and consequently, the array power impose greater voltage transients, while changing the array voltage with ramp yields lower transients and much better power quality. PCC voltage waveforms with different MPPT algorithms are compared in Fig. 10 (e) and (f). High transient voltage caused by the two other algorithms is harmful for the system, especially in micro-grid applications. Efficiency of the methods such as the one presented in [11] depends on uniformity of the array modules. Since sampling from of the array P-V characteristic in these methods are done with specific intervals (e.g. 0.7 $V_{oc-mod}$) which depend on the model of modules, non-identical modules in the array affect their Efficiency. In contrast, the proposed MPPT algorithm is completely independent from the modules make and model. In Table IV, features of the three afore-mentioned methods are compared with each other in different aspects. Superiority of the proposed method over the other methods is obvious from the presented results in Fig. 10 and Table IV.

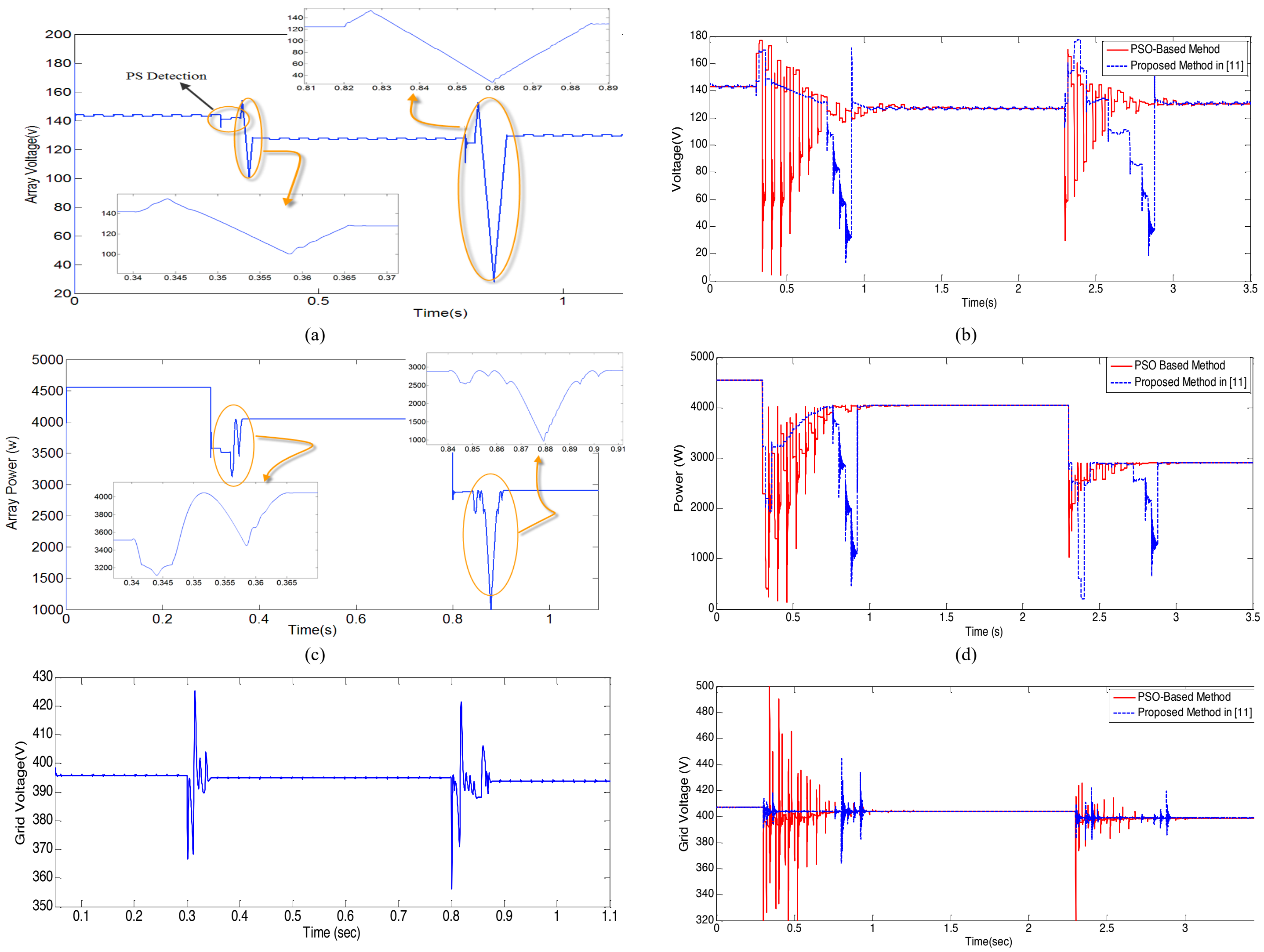

Fig. 10. MPPT process with the proposed and two other methods in different PSC patterns. 1- The proposed method: (a) array voltage, (c) array power, and (e) grid voltage. 2- Two other methods: (b) array voltage, (d) array power, and (f) grid voltage.

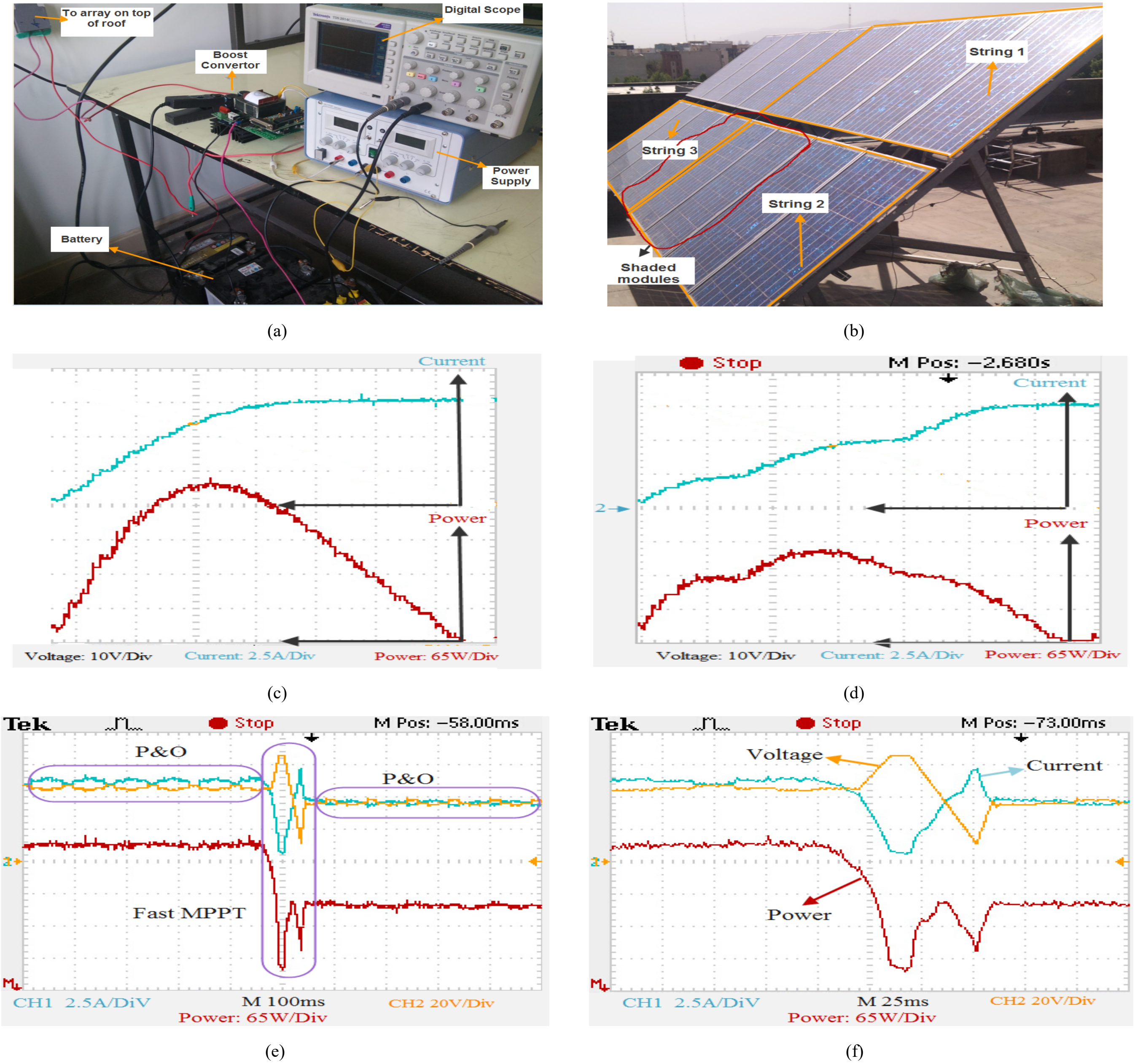

Fig. 11. Overview of experimental setup: a) power stage and b) PV array, I-V and P-V characteristics of tested array c) under UIC and d) under PSC, e-f) MPP tracking response with the proposed method.

TABLE IV: COMPARISON OF THREE MPPT METHODS.

| Criteria | The new proposed method | PSO base method in [19] | Proposed method in [11] |
|---|---|---|---|
| Speed of tracking | 40ms (increasable) | About 500ms | More than 500ms |
| Transients and stress on converter | Low | High | Medium |
| Imposed transient voltages on the grid | Low | High | Medium |
| Energy loss | Low | High | High |
| Computational burden | Low | High | Medium |
| dependency to uniformity of modules | No | A little | Yes |
| Success probability | In all situations | Is not proved in all situations | almost in all situations |

### B. Experimental Results

To verify the performance of the proposed GMPPT algorithm, an experimental setup is developed and the proposed MPPT method is applied to it. Fig. 11 (a) and (b) show the experimental setup. The setup comprises a boost converter that is paralleled with 8 batteries with total 96V to keep its output voltage ($v_o$) constant. It is noteworthy that because $v_o$ is used in determination of desired duty cycle, its transients do not deteriorate efficiency of the proposed method. Input of the boost converter is connected to a PV array with 3 strings; each string consisting of 8 series modules. In standard condition, $V_{oc-mod}$ and $I_{sc}$ are 10V and 2.7A, respectively. In this system, Atmel AVR ATMEGA16A microcontroller is used as the main controller and switching frequency of PWM has been set to 20 kHz.

For testing the performance of the proposed MPPT method, 2 modules from string 2 and 1 module from string 3 are shaded. In 0 Fig. 11 (b) shaded modules of the array are shown. The P-V and I-V characteristics of the array under UIC and this PSC are shown in Fig. 11 (c) and (d). Because of weariness of the PV modules, MPP has occurred in a lower voltage, and efficiency of the array has been decreased. In the given PSC, the system has 3 local maxima.

At first the system operates under UIC and the P&O method carries out MPPT. Suddenly, the PS occurs and the proposed method starts to work and tracks the maximum power point in about 60ms. The current, voltage and power of the array during the process are shown in Fig. 11 (e), and for better observation, their zoomed views are also presented in Fig. 110 (f). It is observed that similar to simulations, GMPP is tracked quickly and smoothly using the proposed algorithm, and after tracking the GMPP, P&O algorithm has been activated again. According to the results in Fig. 10 and Fig. 110, it is obvious that the MPPT process in simulation and experiments are compatible. Saturation of the array voltage that is shown in Fig. 11 (f) is because of the low output voltage of the boost converter.

## VII. Conclusion

In this paper, firstly a partial shading condition detection algorithm is presented and its performance is studied in different situations. The proposed algorithm determines whether the system operates at uniform irradiance or not.

A novel simple and fast algorithm is then presented for MPPT under PSC that operates as direct control method and needs no feedback control of current and voltage. The algorithm is based on ramp change of PV array voltage and simultaneous sampling of its voltage and current continuously. Simulation and experimental results validate the performance of the proposed method in speed and accuracy. The proposed GMPPT method has the following benefits: 1- It is simple and can be implemented with a cheap microcontroller like AVR; 2- It has a high adjustable speed; 3- Because of the smooth change of power in comparison with other methods, it has minimum negative impact on the connected power system; and 4- Its efficiency is guaranteed and is not dependent to the model of modules.

## References

[1] T. Esram and P. L. Chapman, "Comparison of Photovoltaic Array Maximum Power Point Tracking Techniques," *Energy Conversion, IEEE Transactions on,* vol. 22, pp. 439-449, 2007.

[2] Y.-J. Wang and P.-C. Hsu, "An investigation on partial shading of PV modules with different connection configurations of PV cells," *Energy,* vol. 36, pp. 3069-3078, 2011.

[3] D. Kun, B. XinGao, L. HaiHao, and P. Tao, "A MATLAB-Simulink-Based PV Module Model and Its Application Under Conditions of Nonuniform Irradiance," *Energy Conversion, IEEE Transactions on,* vol. 27, pp. 864-872, 2012.

[4] J. Young-Hyok, J. Doo-Yong, K. Jun-Gu, K. Jae-Hyung, L. Tae-Won, and W. Chung-Yuen, "A Real Maximum Power Point Tracking Method for Mismatching Compensation in PV Array Under Partially Shaded Conditions," *Power Electronics, IEEE Transactions on,* vol. 26, pp. 1001-1009, 2011.

[5] E. Koutroulis and F. Blaabjerg, "A New Technique for Tracking the Global Maximum Power Point of PV Arrays Operating Under Partial-Shading Conditions," *Photovoltaics, IEEE Journal of,* vol. 2, pp. 184-190, 2012.

[6] N. Tat Luat and L. Kay-Soon, "A Global Maximum Power Point Tracking Scheme Employing DIRECT Search Algorithm for Photovoltaic Systems," *Industrial Electronics, IEEE Transactions on,* vol. 57, pp. 3456-3467, 2010.

[7] Syafaruddin, E. Karatepe, and T. Hiyama, "Artificial neural network-polar coordinated fuzzy controller based maximum power point tracking control under partially shaded conditions," *Renewable Power Generation, IET,* vol. 3, pp. 239-253, 2009.

[8] I. Abdalla, J. Corda, and L. Zhang, "Multilevel DC-Link Inverter and Control Algorithm to Overcome the PV Partial Shading," *Power Electronics, IEEE Transactions on,* vol. 28, pp. 14-18, 2013.

[9] P. Sharma and V. Agarwal, "Exact Maximum Power Point Tracking of Grid-Connected Partially Shaded PV Source Using Current Compensation Concept," *Power Electronics, IEEE Transactions on,* vol. 29, pp. 4684-4692, 2014.

[10] C. Woei-Luen and T. Chung-Ting, "Optimal Balancing Control for Tracking Theoretical Global MPP of Series PV Modules Subject to Partial Shading," *Industrial Electronics, IEEE Transactions on,* vol. 62, pp. 4837-4848, 2015.

[11] H. Patel and V. Agarwal, "Maximum Power Point Tracking Scheme for PV Systems Operating Under Partially Shaded Conditions," *Industrial Electronics, IEEE Transactions on,* vol. 55, pp. 1689-1698, 2008.

[12] Y. Wang, Y. Li, and X. Ruan, "High Accuracy and Fast Speed MPPT Methods for PV String Under Partially Shaded Conditions," *Industrial Electronics, IEEE Transactions on,* vol. PP, pp. 1-1, 2015.

[13] A. Kouchaki, H. Iman-Eini, and B. Asaei, "A new maximum power point tracking strategy for PV arrays under uniform and non-uniform insolation conditions," *Solar Energy,* vol. 91, pp. 221-232, 2013.

[14] C. Kai, T. Shulin, C. Yuhua, and B. Libing, "An Improved MPPT Controller for Photovoltaic System Under Partial Shading Condition," *Sustainable Energy, IEEE Transactions on,* vol. 5, pp. 978-985, 2014.

[15] T. Kok Soon and S. Mekhilef, "Modified Incremental Conductance Algorithm for Photovoltaic System Under Partial Shading Conditions and Load Variation," *Industrial Electronics, IEEE Transactions on,* vol. 61, pp. 5384-5392, 2014.

[16] L. Yi-Hwa, H. Shyh-Ching, H. Jia-Wei, and L. Wen-Cheng, "A Particle Swarm Optimization-Based Maximum Power Point Tracking Algorithm for PV Systems Operating Under Partially Shaded Conditions," *Energy Conversion, IEEE Transactions on,* vol. 27, pp. 1027-1035, 2012.

[17] K. Ishaque, Z. Salam, M. Amjad, and S. Mekhilef, "An Improved Particle Swarm Optimization (PSO)–Based MPPT for PV With Reduced Steady-State Oscillation," *Power Electronics, IEEE Transactions on,* vol. 27, pp. 3627-3638, 2012.

[18] K. Ishaque and Z. Salam, "A Deterministic Particle Swarm Optimization Maximum Power Point Tracker for Photovoltaic System Under Partial Shading Condition," *Industrial Electronics, IEEE Transactions on,* vol. 60, pp. 3195-3206, 2013.

[19] M. Miyatake, M. veerachary, F. Toriumi, N. Fujii, and H. Ko, "Maximum Power Point Tracking of Multiple Photovoltaic Arrays: A PSO Approach," *Aerospace and Electronic Systems, IEEE Transactions on,* vol. 47, pp. 367-380, 2011.

[20] K. Sundareswaran, S. Peddapati, and S. Palani, "MPPT of PV Systems Under Partial Shaded Conditions Through a Colony of Flashing Fireflies," *Energy Conversion, IEEE Transactions on,* vol. 29, pp. 463-472, 2014.

[21] S. Lyden and E. Haque, "A Simulated Annealing Global Maximum Power Point Tracking Approach for PV Modules under Partial Shading Conditions," *Power Electronics, IEEE Transactions on,* vol. PP, pp. 1-1, 2015.

[22] G. Petrone, G. Spagnuolo, R. Teodorescu, M. Veerachary, and M. Vitelli, "Reliability Issues in Photovoltaic Power Processing Systems," *Industrial Electronics, IEEE Transactions on,* vol. 55, pp. 2569-2580, 2008.

[23] X. Weidong, W. G. Dunford, P. R. Palmer, and A. Capel, "Regulation of Photovoltaic Voltage," *Industrial Electronics, IEEE Transactions on,* vol. 54, pp. 1365-1374, 2007.

[24] S. Moballegh and J. Jiang, "Modeling, prediction, and experimental validations of power peaks of PV arrays under partial shading conditions," *Sustainable Energy, IEEE Transactions on,* vol. 5, pp. 293-300, 2014.

[25] X. Haitao, Y. Yangguang, and J. Haijiang, "A two-stage PV grid-connected inverter with optimized anti-islanding protection method," in *Sustainable Power Generation and Supply, 2009. SUPERGEN '09. International Conference on*, 2009, pp. 1-4.